\documentclass[twocolumn,prb,superscriptaddress]{revtex4}
\usepackage{graphicx}
\usepackage{dcolumn}
\usepackage{bm}
\usepackage{amsmath}
\usepackage{times}
\usepackage{color}
\usepackage[breaklinks=true,colorlinks,citecolor=blue,linkcolor=blue,urlcolor=blue]{hyperref}

\makeatletter

\newcommand{\Rmnum}[1]{\expandafter\@slowromancap\romannumeral #1@}
\makeatother

\begin{document}

\title{Non-Fermi liquid regime in metallic pyrochlore iridates: Quantum Griffiths singularities}
\author{Bikash Ghosh}
\affiliation{Department of Physics, Indian Institute of Technology Kanpur, Kanpur - 208016, India}
\author{Vinod Kumar Dwivedi}
\thanks{Present Address: Department of Physics, Indian Institute of Technology Bombay, Mumbai 400076, India}
\affiliation{Materials Science Programme, Indian Institute of Technology Kanpur, Kanpur - 208016, India}
\author{Soumik Mukhopadhyay}
\email{soumikm@iitk.ac.in}
\affiliation{Department of Physics, Indian Institute of Technology Kanpur, Kanpur - 208016, India}

\begin{abstract}

We investigate the interplay of Kondo and RKKY coupling in presence of disorder by chemically substituting local moment Cr$^{3+}$ at the Ir sublattice in the metallic Pr$_{2}$Ir$_{2}$O$_{7}$. We find evidence of non-Fermi liquid (NFL) behaviour in the transport and thermodynamic measurements at low temperature. Specifically, the magnetic susceptibility exhibits power law divergence at $T=0$. The nonanalytic temperature and magnetic-field dependence of magnetic susceptibility and the associated scaling suggest existence of a two-fluid system consisting of Kondo-screened paramagnetic metal coexisting with magnetically ordered rare regions dominated by inter-impurity interaction, similar to quantum critical Griffiths phase.
\end{abstract}

\maketitle
Pyrochore 5d Iridates R$_{2}$Ir$_{2}$O$_{7}$ (R=Y, or rare earth elements)~\cite{Krempa,Pesin,Gardner,Wan,Machida,Rau,Schaffer,Abhishek1,Abhishek2} offer an ideal template to study the inter-play of spin-orbit coupling (SOC), Coulomb correlation and crystal electric field (CEF) effect in presence of geometric frustration, potentially leading to the emergence of novel topological electronic and magnetic phases such as Weyl semimetal, chiral spin liquid, etc. An interesting member of this family of compounds is Pr$_{2}$Ir$_{2}$O$_{7}$ (PIO), with a non-Kramers doublet ground state for the local moment Pr$^{3+}$. PIO is supposed to be a two channel Kondo system with a rich phase diagram~\cite{Flint,Sung}. However, it was initially proposed that the Kondo coupling is dominated by RKKY interaction within the two-impurity Kondo effect framework and that the long range magnetic order is not suppressed due to Kondo screening, but due to the geometric frustration. This turns PIO into a 'metallic-spin liquid'~\cite{Nakatsuji,Machida1} on a `Kondo lattice' with a partial freezing of spins far below the antiferromagnetic Curie-Weiss temperature~\cite{Nakatsuji}. The angle resolved photo-emission spectroscopy (ARPES) study revealed that PIO had $3$D quadratic band touching at the Brillouin zone center, which opened up the possibility of realization of non-Fermi liquid (NFL) phases by doping, strain or confinement effects~\cite{Takeshi,Ohtsuki,Cheng}. Recently, it was even suggested that the resistivity minimum of slightly doped bi-axially compressed PIO films was a consequence of interplay between decreasing scattering rate as well as carrier density~\cite{Cheng}, thus bringing into question the widely held understanding of PIO as a Kondo lattice system.

Similar to other strongly correlated systems, a direct route to investigating the evolution of the quantum phases in pyrochlore iridates is by chemical substitution. Several such studies have already been reported, for example, in Bi doped  Eu$_{2}$Ir$_{2}$O$_{7}$~\cite{Prachi}; chemical doping of Ca$^{2+}$ for Nd$^{3+}$ in Nd$_{2}$Ir$_{2}$O$_{7}$~\cite{Porter}, etc. In PIO, the local moment comes from Pr$^{3+}$ ion and conduction electrons are mainly contributed by Ir$^{4+}$~\cite{Sung,Ikeda}. The Kondo ($T_{K}$) and RKKY ($T_{RKKY}$) energy scales are very close to each other in PIO~\cite{Nakatsuji}. Chemical substitution in PIO offers us with an opportunity to study the NFL behaviour, if any, arising primarily due to the competition between the intra-site Kondo and the inter-site RKKY interactions in the midst of a disordered environment. The ratio of the two energy scales may be tuned by chemically substituting either the Pr$^{3+}$ or Ir$^{4+}$ site. The altered energy scale is expected to provide driving force for the emergence of novel ground states or perhaps more importantly, to provide a better understanding of the ground state in the un-substituted material. The competition between the intra-site Kondo and the inter-site RKKY interactions within a disordered environment can have profound consequences especially close to the quantum critical point. In the present article, we investigate the evolution of magnetic and electronic properties by gradual substitution of Cr$^{3+}$ for Ir$^{4+}$ in PIO. Since, Cr$^{3+}$ is a 3d magnetic ion, the substitution will bring an extra local moment impurity in the Ir sublattice, besides already existing Pr$^{3+}$~\cite{Ikeda}. Both Kondo and RKKY energy scales are sensitive functions of carrier density~\cite{Doniach}: the substitution of Cr$^{3+}$ for Ir$^{4+}$ bring changes in the carrier concentration of the system which in turn affect the f-d exchange interaction among Ir-5d and Pr-4f localized spins. Moreover, Cr$^{3+}$, being lighter than Ir$^{+4}$, is likely to introduce disorder by modifying Ir-O bond length and Ir-O-Ir bond angle.


Bulk polycrystalline samples of Pr$_2$Ir$_{2-x}$Cr$_x$O$_7$ (PICO), where $x=0, 0.05, 0.1, 0.2, 0.3$, were grown by conventional solid state reaction route~\cite{Vinod1,Vinod2,Matsuhira,Vinod3} using powders of Pr$_2$O$_3$, IrO$_2$ and Cr$_2$O$_3$, mixed in their stoichiometric ratios. The samples were characterized by powder x-ray diffraction (XRD) measurement using PANalytical XPertPro diffractometer at room temperature. The electrical transport measurement was performed using conventional four probe technique. Magnetic susceptibilities were measured by a Quantum Design physical property measurement system (PPMS). The electronic states were characterized by X-ray photo-emission spectroscopy (XPS) using PHI 5000 Versa Probe II system.

\begin{figure}
	\includegraphics[width=\linewidth]{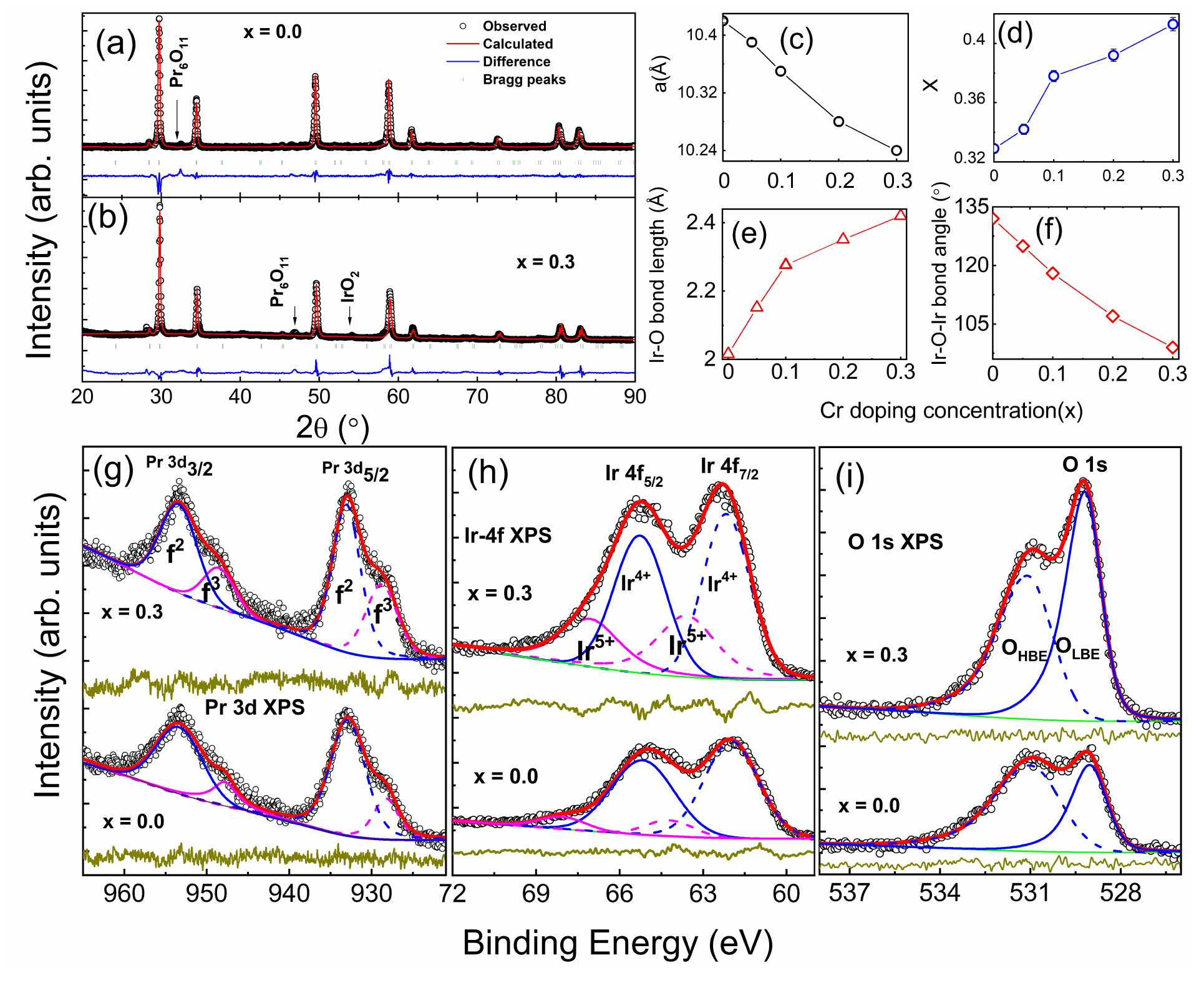}
	\caption{XRD spectra along with Rietveld refinement recorded at room temperature for Pr$_2$Ir$_{2-x}$Cr$_x$O$_7$ with (a) $x=0.0$ and (b) $x=0.3$. Arrows indicate nominal presence of parasitic phases. (c) Variation of lattice parameter $a$ as a function of $x$ (d) Variation of positional parameter $X$ as a function of $x$. Variation of (e) $Ir-O$ bond distance, and (f) $Ir-O-Ir$ bond angle against $x$. The XPS spectra of (g) $Pr-3d$ for $x=0.0$ and $0.3$, (h) $Ir-4f$ spectra for $x=0.0$ and $x=0.3$ samples, and (i) De-convoluted $O-1s$ peaks for $x=0.0$ and $x=0.3$ samples.}\label{fig:xrd}
\end{figure}

Fig.~\ref{fig:xrd}a, b shows the powder XRD pattern taken at room temperature along with Rietveld refinement using FullProf suite for two representative PICO samples with $x=0.0$ and $x=0.3$, respectively. The refinement shows cubic fcc pyrochlore phase with space group Fd$\bar{3}m$ with no notable changes in terms of emergence of new peaks or changes in peak position with chemical substitution. The goodness of fit ($GOF=\frac{R_{wp}}{R_{exp}}$, where $R_{wp}$ and $R_{exp}$ is the expected weighed profile factor and the observed weighed profile factor, respectively) value ranges between $2.3-8$ for all samples. We observe variation in structural parameters with chemical substitution as well. It is expected that mismatch in the ionic radii between Cr$^{3+}$ ($\sim 0.615$\AA) and Ir$^{4+}$ ($\sim 0.625$ \AA) would lead to changes in structural parameters. The lattice parameters obtained from Rietveld refinement using FullProf suite are plotted in Fig.~\ref{fig:xrd}c-f. The lattice constant $a$ for $x=0.0$ turns out to be $10.42$\AA, which is consistent with the value reported in literature~\cite{Millican,Takatsu}. It decreases with doping of $Cr$ content $x$ shown in Fig.~\ref{fig:xrd}c. In Pyrochlore structure, the variable positional parameter $X$ plays a crucial role in controlling the magnetic and electrical transport properties~\cite{Krempa,Vinod1}. The variation of positional parameter $X$ as a function of $Cr$ doping content $x$ is shown in Fig.~\ref{fig:xrd}d. The value of $X$ for an ideal $IrO_6$ octahedra in Pyrochlore structure is $0.3125$, where the $Ir^{4+}$ ions are placed under an ideal cubic symmetry~\cite{Gardner,Vinod1}. The estimated value for parent sample $x=0.0$ turns out to be $0.329$, which is close to the ideal value. With substitution, the $X$ value is enhanced, indicating more distortion in Ir$O_6$ octahedra. The resultant change in crystal electric field alters the local hybridization between $Ir(5d)/Cr(3d)$ and $O(2p)$, $Pr(4f)$ states. The variation of $Ir-O$ bond length and $Ir-O-Ir$ bond angle as a function of $x$ is shown in Fig.~\ref{fig:xrd})e,f, respectively. The $Ir-O$ bond length increases with $x$, while on the other hand, the $Ir-O-Ir$ bond angle is reduced with increasing substitution.


Fig.~\ref{fig:xrd}g-i shows XPS spectra taken at room temperature for the two representative samples $x=0.0$, and $0.3$ of PICO. We fitted the spectrum of $Pr-3d$ XPS using a Tougaard-type background~\cite{Gamza}. For $x=0$, fitting suggests a majority of Pr$^{3+}$ charge state. The $Pr-3d$ states are split into two peaks, i.e., $Pr-3d_{5/2}$ and $Pr-3d_{3/2}$ centered around binding energy $933$ eV and $953$ eV, respectively. The two peaks located at binding energies $933$ eV (main peak) and $928$ eV (satellite peak) are labeled as $f^2$ and $f^3$ states, respectively~\cite{Lutkehoff,Ishii}. The ratio of two peaks, i.e., $I_{928}$ (satellite peak)/$I_{933}$(main peak) turns out to be $0.26$ ($x=0.0$) and $0.5$ ($x = 0.3$) for the Pr$^{3+}$ ion, suggesting sizable fraction of $Pr^{+4}$ charge states along with Pr$^{3+}$~\cite{Lutkehoff}. A rough estimation of the coupling between the conduction electrons and the $Pr-4f$ orbital ($\bigtriangleup$) is given by the intensity ratio $r=I(f^3)/[I(f^2)+I(f^3)]$~\cite{Gamza,Ishii,Slebarski}. The value of $r$ ($0.21$ for $x=0.0$, and $0.35$ for $x=0.3$) enhances with $Cr$ substitution suggesting increased hybridization between the conduction electrons and the $Pr-4f$ orbital. The strong mixing of conduction electrons and the $Pr-4f$ orbital and the evidence of charge fluctuation as suggested by the existence of Pr$^{4+}$, are characteristic of Kondo systems ~\cite{Gamza,Ishii,Slebarski}. The de-convoluted $Ir-4f$ XPS spectra of $x=0.0$ and $x=0.3$ samples are shown in Fig.~\ref{fig:xrd}h. We find that Ir$^{4+}$ charge state is in majority, while $Ir^{5+}$ is present nominally ($\sim 7\%$) in parent sample ($x=0$). The Ir$^{5+}$ component is increased with substitution ($30\%$ for $x=0.3$). Fig.~\ref{fig:xrd}i shows the $O-1s$ spectra with two clearly resolved peaks. The peak situated at lower binding energy $529$ eV is labeled as O$_{LBE}$ and the peak at the higher binding energy $531$ eV, as O$_{HBE}$. The O$_{LBE}$ peak is due to O$^{-2}$ anion in the system and the O$_{HBE}$ peak is possibly associated with the presence of O-Pr$^{4+}$~\cite{Novica}.
\begin{figure*}
	\includegraphics[width=12 cm]{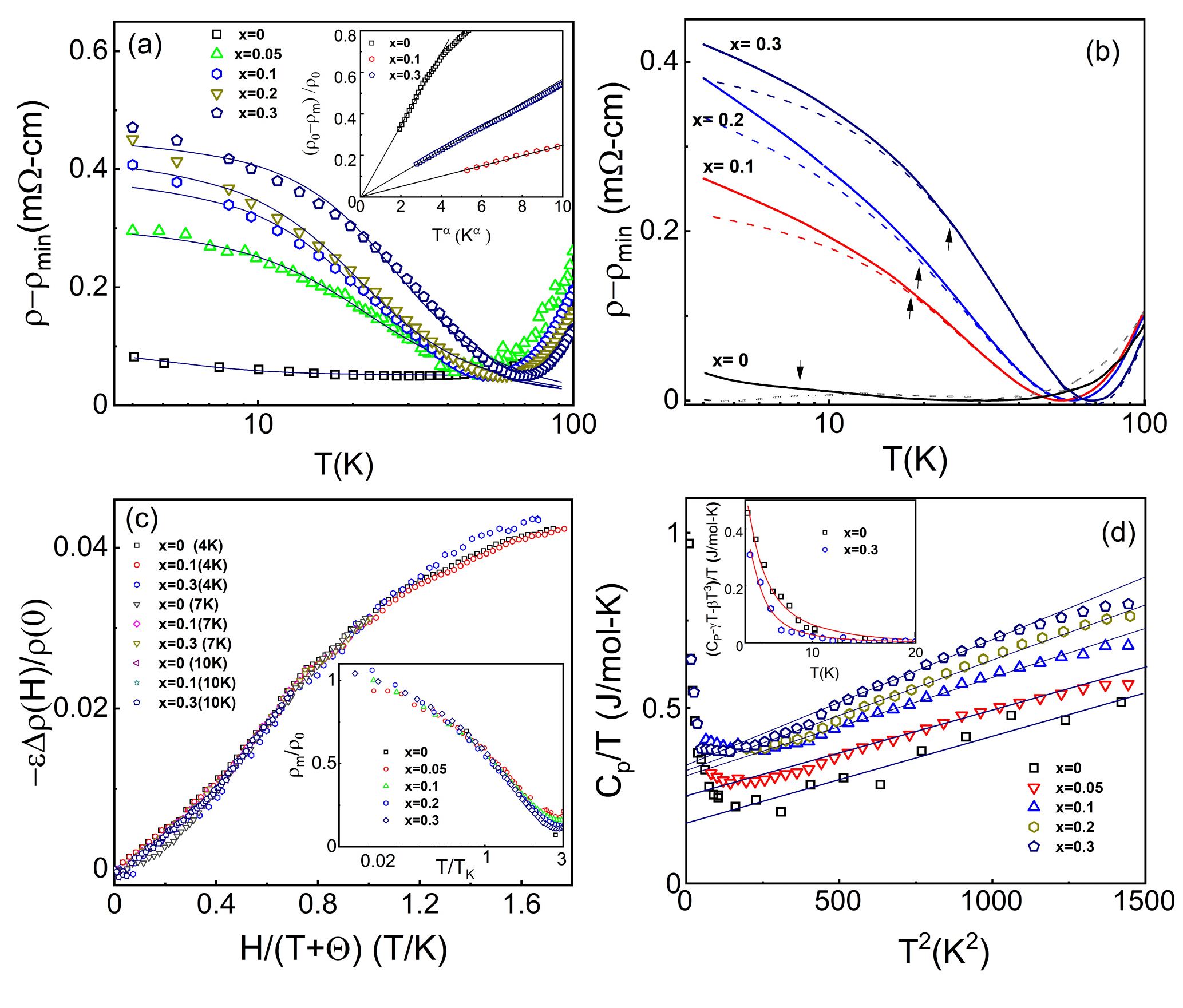}
	\caption{(a) The temperature dependence of spin contribution to the resistivity ($\rho_{m}$) of PICO in the absence of magnetic field, where $\rho_{m}$ is calculated by subtracting resistivity minima $\rho_{min}$ from $\rho(T)$ at low T, i.e. $\rho_{m}=\rho(T)-\rho_{min}$. The Kondo temperature is estimated using Hamann's expression which is shown by the solid line. Inset: enlarged view of low $T$ resistivity showing power law divergence. (b) The temperature dependence of resistivity in absence and in presence of magnetic field showing suppression of resistivity upturn with magnetic field. The dashed lines correspond to magnetic field $5$~T. The locations of $T_K$ is shown by arrowheads. (c) The magnetoresistance (MR) as a function of $H/T$ for the three representative samples showing universal scaling. Inset: The magnetic resistivity $\rho_{m}$ divided by residual resistivity $\rho_{0}$ plotted against normalized temperature $T/T_{K}$ for various $x$, are found to overlap with each other, demonstrating the existence of a universal scaling. (d) Specific heat divided by temperature $C_{P}/T$ is plotted against $T^{2}$ for various x. The straight lines passing through the data points are linear fits with $\gamma + \beta T^{2}$. Inset: The NFL component at low temperature plotted for $x=0$ and $x=0.3$. The continuous lines are guides to the eye.}\label{fig:RT}
\end{figure*}

Fig.~\ref{fig:RT}a shows magnetic contribution to the resistivity $\rho_{m}$ in absence of magnetic field. There is a rapid increase of $\rho_{m}$ with increasing $x$ with the ratio $(\Delta\rho)_{max}/\rho_0$ ($\rho_0$ being the residual resistivity) remaining unchanged, which suggests dominance of single impurity scattering in transport. It is also observed that the resistivity minima are shifted towards higher temperature with increasing substitution implying higher Kondo temperature $T_K$. In general, the magnetic resistivity follows the generalized Hamann's expression describing potential scattering at each impurity site~\cite{Hamann, Fischer}. A clear tendency towards saturation of resistivity at low temperature develops with substitution. The Kondo temperature $T_{K}$, extracted from Hamann's fit, increases with increasing substitution, from $7$ K for $x=0$ to $24$ K for $x=0.3$. We calculated residual resistivity $\rho_{0}$ by extrapolating Hamann's fit to $T=0$, which are found to be increasing with $x$. The rapidly rising $\rho_{0}$ is expected and it is mainly due to increasing $Ir$ substitution by magnetic $Cr$. At still lower temperature, we observe deviation from Hamann's fit below around $6$~K. In fact, the resistivity seems to follow power law at very low temperature as shown in the inset of Fig.~\ref{fig:RT}a with the exponent ranging between $0.7-1.2$, suggesting non-Fermi-liquid behaviour. However, the temperature range of the deviation from $T^2$ behaviour is too small so as to arrive at a definite conclusion. For example, the residual resistivity $\rho_{0P}$ estimated from the power law fit is only slightly higher than that estimated from Hamann's fit, $\rho_{0H}$ (Table~\ref{T1}).

In Fig.~\ref{fig:RT}b, we show the temperature dependence of resistivity at low temperature in presence of magnetic field. For $x=0$, the resistivity upturn is almost completely suppressed leading to negative magnetoresistane (MR). For $x\neq0$, the MR becomes negligible roughly above the Kondo temperature and increases with decrease of temperature. Appearance of negative MR at temperatures far below $T_K$ could be due to contribution by NFL component at low temperature. The maximum value of negative MR observed at the lowest measured temperature is actually reduced marginally with substitution (Table~\ref{T1}, Fig.~\ref{fig4}b). It turns out that the MR is a universal function of $H/T$ for all $x$ (Fig.~\ref{fig:RT}c), which is again a characteristic of Kondo systems~\cite{Daybell,singh}. The parameter $\Theta$ used in MR scaling has the value in the range $0.1-0.2$~K. Another standard experimental signature of Kondo screening effect is the universal scaling of Kondo resistivity~\cite{singh,Daybell,Schilling,Anderson}. Inset of Fig.~\ref{fig:RT}c shows $\rho_{m}/\rho_{0}$ as a function of $T/T_{K}$ for all $x$, overlapping with each other below the resistivity minima.


We performed specific heat ($C_{p}$) measurement at low temperature. The specific heat increases as a function of Cr concentration. In the temperature range $15-35$ K, $C_{p}/T$ varies linearly with $T^{2}$ as shown in the Fig.~\ref{fig:RT}d. The solid lines are the linear fits with $\gamma+\beta T^{2}$, where $\gamma$ is the linear coefficient. The linear coefficient $\gamma$ is determined from the intercept of the linear fit with vertical axis at $T=0$. The extracted $\gamma$ value from the intercepts increases with increase of $x$ as shown in Fig.~\ref{fig:RT}d and Table~\ref{T1}. The $\gamma$ value is about $500$ times larger in $x=0.3$ than copper. The high value of linear coefficient for all samples confirms quasiparticles gaining heavy thermodynamic mass with Cr substitution. The second term in the linear fit is due to phonon contribution to the specific heat with $\beta$ being a constant related to the Debye temperature (Table~\ref{T1}). Interestingly, at low temperature $C_P/T$ increases sharply with lowering of temperature, again suggesting NFL behaviour at low temperature. The non-Fermi liquid component is estimated by the total specific heat minus the linear $T$ and $T^3$ component as plotted in the inset of Fig.~\ref{fig:RT}d. The NFL component is marginally reduced with substitution as compared to $x=0$ (Inset, Fig.~\ref{fig:RT}d).

\begin{figure}
	\includegraphics[width=\linewidth]{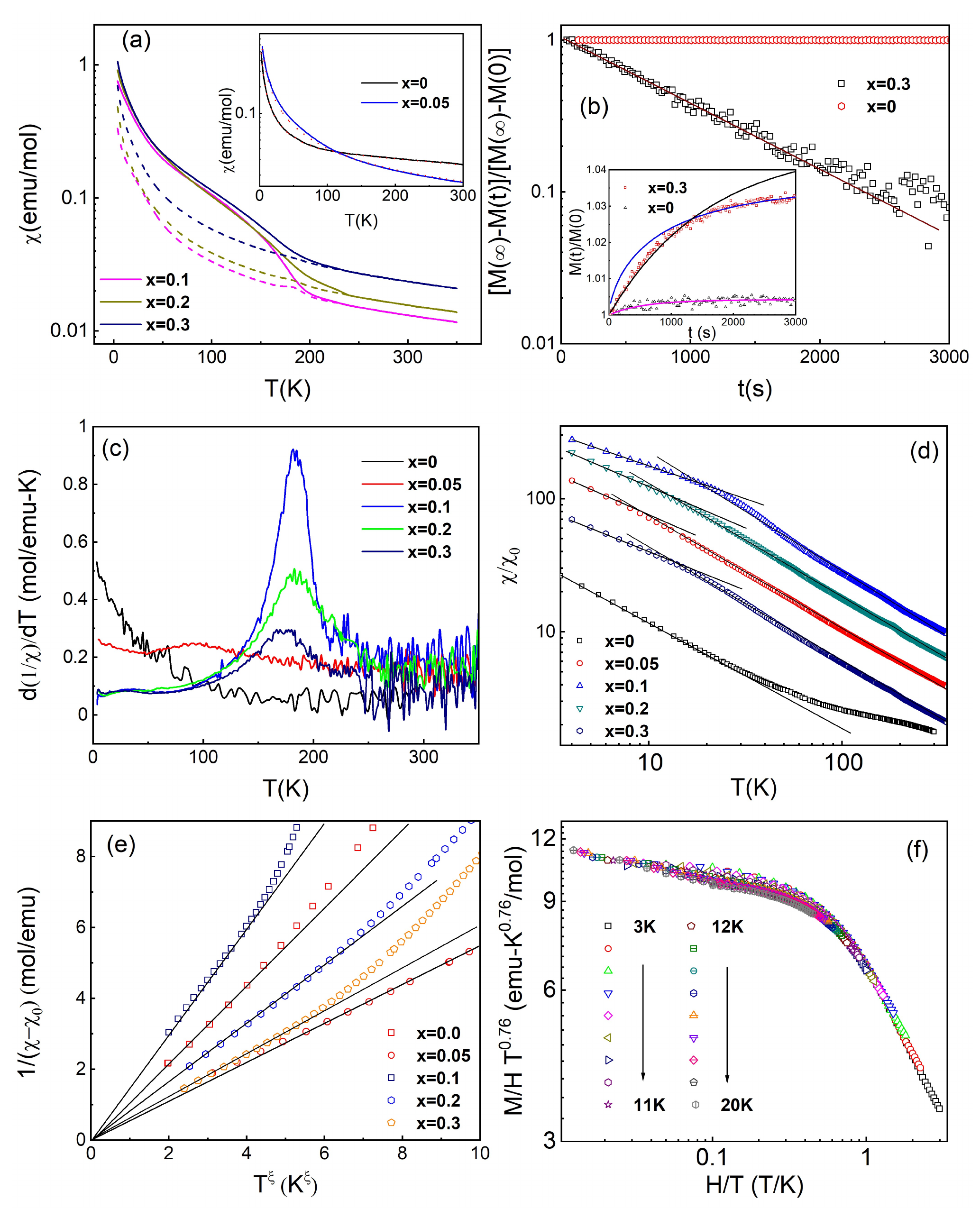}
	\caption{(a) Semi-log plot of dc magnetic susceptibility $\chi(T)$ of Pr$_2$Ir$_{2-x}$Cr$_x$O$_7$ ($x=0.1, 0.2, 0.3$), showing irreversibility in field cooled (FC) and zero field cooled (ZFC) susceptibility. The dashed line represents ZFC $\chi(T)$ while FC $\chi(T)$ is represented by solid line. Inset: $\chi(T)$ for $x=0, 0.05$ show absence of bifurcation in ZFC and FC data. (b) Time dependence of normalized magnetization [M($\infty$)-M(t)]/[M($\infty$)-M(0)] for the two representative samples $x=0.0$ and $x=0.3$ at $5$~K showing pure exponential relaxation at lower time scales in the semi-log plot. Inset: The normalized magnetization is fitted with a pure exponential function at shorter timescale and a stretched exponential function at longer time scale. The extrapolated fits in the two regions are shown by continuous lines. (c) Temperature derivative of inverse FC susceptibility showing evolution of the maximum with substitution. (d) Log-log plot of $\chi/\chi_0$ as a function of temperature showing power law divergence of susceptibility. (e) $1/\chi-\chi_0$ tends to zero as $T$ approaches zero. (f) The scaling plot of $(M/H)T^{\xi_L}$ against $H/T$ for $x=0$.}\label{fig:MT}
\end{figure}

The temperature dependence of the field cooled (FC) and zero field cooled (ZFC) magnetic susceptibility is shown in Fig.~\ref{fig:MT}a. For the parent compound ($x=0$) as well as for $x=0.05$, there is no appreciable bifurcation between FC and ZFC susceptibilities down to the lowest measured temperature $3$~K as shown in the inset of Fig.~\ref{fig:MT}a. As the substitution is increased further, there is clear evidence of opening up of irreversibility in the FC and ZFC curves below certain temperature $T_{irr}$ (Table~\ref{T1}) without any peak in the ZFC curves. The bifurcation is not as pronounced as observed in spin glass systems (That is why a semi-log plot is used). We analyzed magnetic susceptibility of the parent compound ($x=0$) at different temperature regime using the Curie-Weiss (CW) law: $\chi(T)=\chi_0+ C/(T-\theta_{CW})$, where C and $\theta_{CW}$ are curie constant and Weiss temperature, respectively. The temperature independent term $\chi_0$ is attributed to the Van Vleck and Pauli paramagnetic contribution due to Ir 5d conduction electrons. For $x=0$, the high temperature effective local magnetic moment turns out to be $\mu_{eff}=3.70\mu_{B}/$Pr, which is very close to the effective moment of $3.58$ $\mu_{B}/$Pr expected for isolated $Pr^{+3}$ ions. The negative $\theta_{CW}\sim -19K$) arises due to effective antiferromagnetic RKKY interaction among the $Pr-4f$ moments. Again, $\mu_{eff}$ and $\theta_{CW}$ are determined by CW fit in the region $2.5-20$~K, which gives diminished values $2.6 \mu_{B}$, $-0.8$~K, respectively, indicating Kondo screening of $Pr-4f$ moments. The $\mu_{eff}$ and $\theta_{CW}$ for the samples with $x\neq0$ are calculated by CW fit in the high temperature regime. It is found that $\theta_{CW}$ is higher for $x\neq0$ compared to that for $x=0$. It is also observed that $\mu_{eff}$ increases with Cr substitution (Table~\ref{T1}). This enhancement in $\mu_{eff}$ is expected as Cr$^{3+}$ possesses large (local) moment ($\mu_{Cr^{3+}}=3.9\mu_{B}$).

\begin{table*}
	\centering
	\caption{Important physical parameters obtained from the magnetization, transport and specific heat measurements}\label{T1}
	\begin{tabular}{c c c c c c c c c c c c c c c c c c}
		\hline
		$x$ &$-\theta_{CW}$ & $\mu_{eff}^{exp}$& $\mu_{eff}^{theo}$ & $\xi_L$ & $\xi_H$ & $\alpha$ & $T_{irr}$ & $\chi_0$ & $T^\ast$ & $T_K$ & $\rho_{0H}$ & $\rho_{0P}$ & $-\frac{\Delta\rho_{max}(7T)}{\rho(0)}$ & $\gamma$ & $\beta$\\
         & (K) & ($\mu_B$) & ($\mu_B$) &   &   &   & (K) & (emu/mol) & (K) & (K) & ($m\Omega cm$) & ($m\Omega cm$) & (@$4$K) & (mJ/mol K) & 10$^4.$J/mol-K$^3$ \\
		\hline
		0.0 & ~19 & ~5.38 & ~5.06 & ~0.74(4) & - & 0.7 & - & 0.015 & - & ~7 & 0.03 & 0.04 & 0.04 & 165  &  1.67\\
	
		0.05 & ~27 & ~6.04 & ~5.14 & ~0.70(6) & ~0.82(1) ~& 0.7 &  - & 0.004 & ~95 & ~18 & 0.32  &  0.36 & 0.04
 & 236  & 2.16\\
	
		0.1 & ~33 & ~5.44 & ~5.21 & ~0.50(4) & ~0.88(2) ~ & ~1.2 & ~235 & 0.001 & ~184 & ~19 & 0.37  & 0.41 &  0.02 & 300 & 2.40\\
	
		0.2 & ~40 & ~6.06 & ~5.35 & ~0.67(5) & ~0.84(1) ~& ~0.9 & ~238 & 0.002 & ~185 & ~20 & 0.41 & 0.47 & 0.04 & 310 & 3.04\\
	
		0.3 & ~45 & ~6.07 & ~5.50 & ~0.63(5) & ~0.85(1) ~& ~0.7 & ~236 & 0.011 & ~172 & ~24 & 0.45 & 0.51 & 0.02 & 332 & 3.25\\
		\hline
	\end{tabular}
\end{table*}

Given the irreversibility in substituted compounds, it was necessary to measure isothermal magnetic relaxation at low temperature. The samples were zero field cooled and after reaching the desired temperature, a dc magnetic field of $0.1$~T was switched on and evolution of magnetization was measured as function of time. Fig.~\ref{fig:MT}b shows time dependence of normalized magnetization for a representative substituted compound $x=0.3$ compared with that for $x=0.0$. After $3000$ sec, the percentage change in $M(t)/M(0)$ is $0.4\%$ for $x=0$ while the same for $x=0.3$ is $3.2\%$ for $=0.3$. Long time magnetic relaxation due to spin glass or Griffiths phase is generally described by a stretched exponential function $M(t)=M(0)+[(M(\infty)- M(0)][1-\exp\{-(t/\tau)^{\beta}\}]$~\cite{Vinod1}, where $\tau$ is characteristic relaxation time and $\beta$ is the stretching exponent which determines the shape of magnetic relaxation. We observe a pure exponential time relaxation \emph{at shorter time scale} with the value of $\beta$ being close to unity for $x=0$ and $0.3$, as demonstrated in the semi-log plot (Fig.~\ref{fig:MT}b). Such a time relaxation seems to suggest formation of magnetically ordered regions with strong spin fluctuation at low temperature. For substituted samples, the slope of the straight line is reduced considerably with increasing temperature and becomes negligible above $15$~K (not shown in the figure), suggesting influence of thermal activation as well. The deviation from pure exponential behavior is not clearly discernible from the semi-log plot. However, on closer inspection, a deviation from pure exponential behavior at longer time scale is indeed observed from the fit of normalized relaxation data in the linear scale (Inset, Fig.~\ref{fig:MT}b). We have introduced piece-wise fitting of the data using pure exponential function at lower time scale and stretched exponential function at larger time scale, with subsequent extrapolations (shown by the continuous lines in the inset of Fig.~\ref{fig:MT}b). Above $1300$s, the $M-t$ data can be best fitted with a stretching exponent ($\beta$)) value of $0.65$ for $x=0.3$. Using the same method, we found non-exponential relaxation with the value of stretching exponent $\beta=0.70$ at $10$ K (not shown in the figure). Such non-exponential relaxation at long timescale could be attributed to the spin glass or Griffiths phase. The temperature derivative of inverse field cooled susceptibility is plotted in Fig.~\ref{fig:MT}c. A well defined extremum develops for $x\geq 0.1$ suggesting magnetic order. In fact, the evolution of the extremum is not abrupt as a faint broad extremum is also observed for $x=0.05$ (Fig.~\ref{fig:MT}c). These extremum points are labeled as $T^\ast$ and are listed in Table~\ref{T1}.
\begin{figure}
\includegraphics[width=\linewidth]{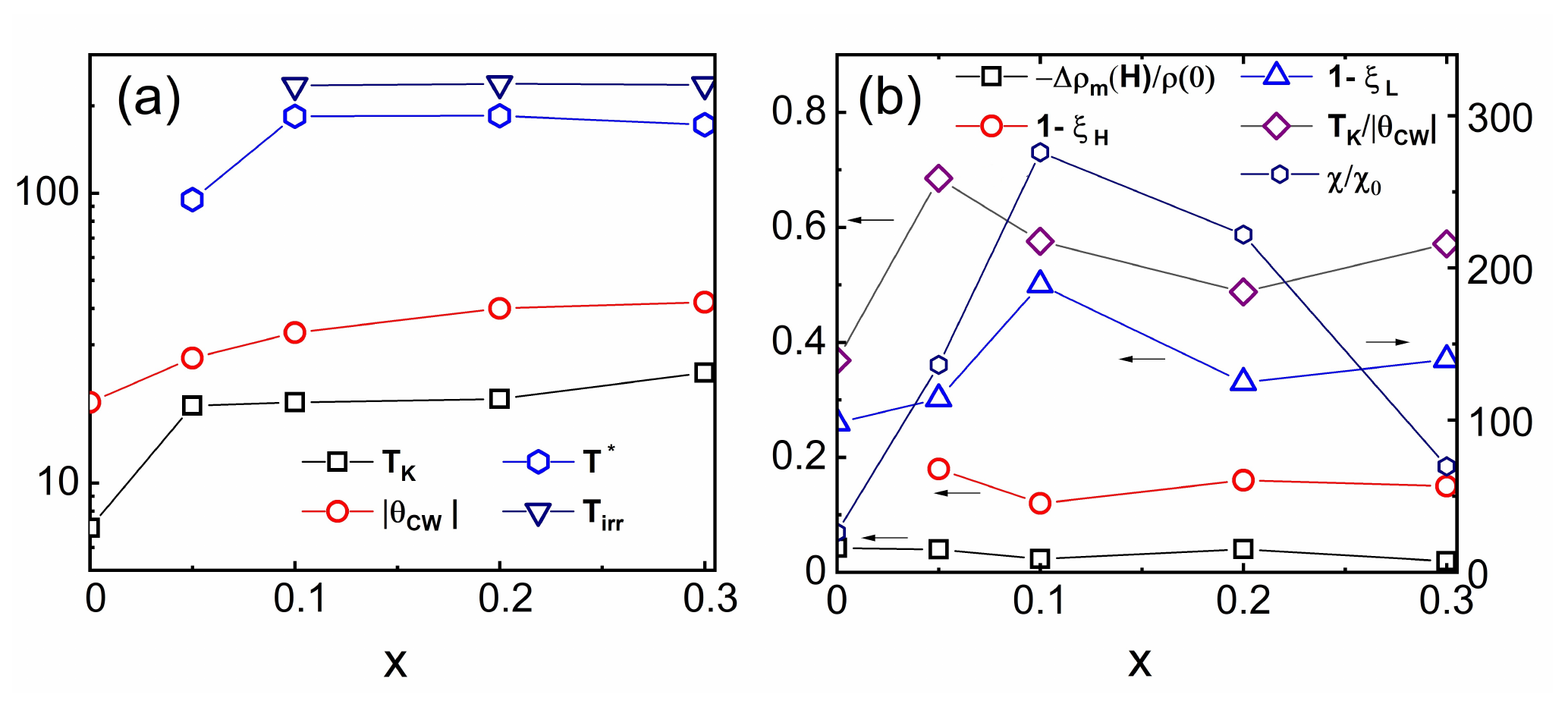}
\caption{(a) Variation of important temperature scales $T_K$, $|\theta_{CW}|$, $T^\ast$ and $T_{irr}$ with $x$. (b) The dimensionless quantities such as the power law exponents $\xi_L$ and $\xi_H$, $\frac{\chi}{\chi_0}$ (at $4$~K), $\frac{T_K}{|\theta_{CW}|}$ and the maximum value of negaive MR plotted against $x$.}\label{fig4}
\end{figure}

Fig.~\ref{fig:MT}d shows log-log plot of ZFC $\chi/\chi_0$ vs $T$ for all samples. As discussed earlier, the value of $\chi_0$ has been determined by CW fit in the paramagnetic regime at high temperature. The susceptibility follows power law divergence $\chi \sim T^{-\xi}$. From the figure, it is clear that at low temperature, the power law exponent is sensitive to $x$ while on the other hand, at high temperature, the variation of $\xi$ with $x$ is practically minimal. Interestingly, for $x=0$, the power law behaviour is limited to low temperature. For nonzero $x$, we observe two power law exponents: one at low temperature ($\xi_L$) and the other at high temperature ($\xi_H$). The exponents are summarized in Table ~\ref{T1}. One of the possible reasons of the difference in low temperature and high temperature power law exponents for $x\neq0$ (with $\xi_L$ being always smaller than $\xi_H$) might be increased Kondo screening of the local moments.

Power law divergence can arise due to enhanced spin fluctuation~\cite{Moriya}, critical valence fluctuations~\cite{Watanabe} or due to heavy quasi-particles coupling with the spin fluctuation in the vicinity of a local quantum critical point (QCP) and getting destroyed in the process~\cite{Coleman}. However, in this case, the magnetic instability, as identified by the irreversibility in the ZFC and FC susceptibility, the extremum in the temperature derivative of FC susceptibility, and power law divergence of ZFC susceptibility, sets in at a much higher temperature scales compared to the respective Kondo temperatures for all samples (table~\ref{T1}). The magnetic instability is thus not emerging from the heavy Fermi liquid. It is not associated with a breakdown of Kondo effect either as the ratio $\chi/\chi_0$ is eventually reduced with increasing substitution (Fig.~\ref{fig4}b, Table~\ref{T1}).

We also test the divergence of $\chi$ at $T=0$ by plotting the inverse of $\chi-\chi_0$ against $T^\xi$ at low temperature and extrapolating the same to $T=0$ (Fig.~\ref{fig:MT}e). We find no finite intercept on the $T=0$ axis for all $x$, thus confirming that the susceptibility truly diverges at $T=0$, going by the trend down to $3$~K. The magnetization at non-zero temperature ($T$) and magnetic field ($H$) exhibits a scaling behavior of $(M/H)T^{\xi_L}$ as a universal function of $H/T^\beta$, characteristic of NFL systems~\cite{Stewart, Adroja, Andraka0}. The $H/T^\beta$ scaling is observed over several decades as shown for $x=0$ in Fig.~\ref{fig:MT}f. For $x=0$, the best collapse of data points is observed in the range $\beta=1\pm 0.1$ and $\xi_L=0.76\pm0.03$. A change in $\beta > 0.1$ and $\xi_{L} > 0.04$ in either direction results in a larger spread of data points than what is plotted in the Fig~\ref{fig:MT}f. Power law divergence of magnetic susceptibility $\chi$, specific heat $C$, and resistivity $\rho$ is observed in case of single impurity fluctuations in a multi channel Kondo system where the power law exponent is associated with conduction electron channel number~\cite{Affleck}. The power law scaling can also be attributed to Griffiths singularities arising out of interplay of between RKKY and Kondo interaction in a disordered Kondo lattice, close to quantum critical point~\cite{Castro}. The competition between RKKY interaction, which favours long range order, and Kondo effect, which leads to quenching of magnetism or local Fermi liquid, in presence of disorder, could lead to magnetically ordered rare regions in the metallic paramagnetic matrix, similar to the Griffiths phase in dilute magnetic systems. The scaling dimension $\beta$ is a reliable indicator of the origin of NFL behaviour. If the value of $\beta$ is greater than unity, the scaling suggests inter-impurity interaction near the quantum critical point whereas $\beta$ value of less than unity suggests single impurity multichannel effect~\cite{Andraka}. In this case, it is difficult to rule, based on the scaling alone, either in favour of single impurity multichannel effect or inter-impurity interaction, as $\beta$ is centered around $1$ with an uncertainty of $\pm 0.1$. However, fortunately, at least the sample with $x=0$ is only capable of exhibiting single impurity two channel Kondo effect due to mixing between $Pr^{3+}$ non-Kramers doublet ground state and Kramers doublet excited state. A two-channel Kondo effect is supposed to show logarithmic divergence in susceptibility and not power law divergence. So we can safely conclude that even for $x=0$, inter-impurity interaction plays a dominant role leading to power law divergence of susceptibility \emph{at low temperature}. For $x\neq0$, the substitution of $Ir$ site by Cr$^{3+}$ increases the local moment concentration which leads to enhanced influence of inter-impurity interaction. This is reflected in the power law behaviour in magnetic susceptibility extending to higher temperature as compared to $x=0$.

We plot the variation of relevant temperature scales and certain dimensionless quantities with substitution $x$ in Fig.~\ref{fig4}a and Fig.~\ref{fig4}b, respectively. We recall from Fig.~\ref{fig:xrd}c that the unit cell volume decreases with increasing $x$. Surprisingly, we observe that that both $T_K$ and $|\theta_{CW}|$ which sets the energy scale for RKKY interaction, are enhanced with substitution. The relatively higher percentage increase of $T_K$ as compared to the increase of $T_{RKKY}$ with chemical substitution (Fig.~\ref{fig4}b) should, on the face of it, inhibit magnetic order due to enhanced quantum fluctuation. The resistivity shows increased tendency towards saturation at low temperature with substitution while the NFL behaviour observed in the un-substituted sample at low temperature is only partially suppressed so far as electrical transport properties and specific heat are concerned. The power law divergence sets in at a much higher temperature compared to $T_K$, suggesting a two fluid system with magnetically ordered rare regions coexisting with Kondo-screened metallic paramagnetic regions. While the macroscopic magnetic susceptibilities are more sensitive to Griffiths singularities arising out of inter-impurity interaction, the resistivity and specific heat are dominated by local Fermi liquid components, except for at very low temperatures.

To conclude, we observe universal scaling of Kondo resistivity as well as magnetoresistance along with enhancement of Kondo and RKKY coupling with local moment Cr$^{3+}$ substitution at the Ir sublattice. However, both Kondo resistivity and specific heat show deviation from Fermi liquid behaviour at low temperature. The low-temperature magnetic susceptibility for $0\leq x \leq 0.3$ is not described as a local Fermi liquid but characterized by a power law divergent behavior, where $\chi \propto T^{-\xi_L}$, with the exponent $0.50\leq \xi_L \leq 0.76$ being significantly less than unity. The magnetization at finite temperature ($T$) and nonzero magnetic field ($H$) exhibits a scaling of $(M/H)T^{\xi_L}$ as a universal function of $H/T^\beta$. The susceptibility at high temperature for the doped samples too follow power law divergence with the exponent varying roughly between $0.8\leq \xi_H \leq 0.9$. The observed non-Fermi liquid behaviour is attributed to magnetically ordered rare regions or the so called Griffiths singularities, close to the quantum critical point.

We acknowledge Department of Science and Technology (DST), government of India for financial support.

\end{document}